\documentclass[11pt,twoside]{article}
\usepackage{asp2010}

\resetcounters

\begin{document}

\title{Fitting Galaxies on GPUs}
\author{Benjamin~R~Barsdell$^1$, David~G~Barnes$^1$, and
  Christopher~J~Fluke$^1$ 
  \affil{$^1$Swinburne University of Technology \\ PO Box 218
    \\ Hawthorn VIC 3122 (Mail H39) \\ Australia} 
}

\begin{abstract}
Structural parameters are normally extracted from observed galaxies by
fitting analytic light profiles to the observations. Obtaining
accurate fits to high-resolution images is a computationally expensive
task, requiring many model evaluations and convolutions with the
imaging point spread function. While these algorithms contain high
degrees of parallelism, current implementations do not exploit this
property. With ever-growing volumes of observational data, an
inability to make use of advances in computing power can act as a
constraint on scientific outcomes. This is the motivation behind our
work, which aims to implement the model-fitting procedure on a
graphics processing unit (GPU). We begin by analysing the algorithms
involved in model evaluation with respect to their suitability for
modern many-core computing architectures like GPUs, finding them to be
well-placed to take advantage of the high memory bandwidth offered by
this hardware. Following our analysis, we briefly describe a
preliminary implementation of the model fitting procedure using
freely-available GPU libraries. Early results suggest a speed-up of
around $10\times$ over a CPU implementation. We discuss the
opportunities such a speed-up could provide, including the ability to
use more computationally expensive but better-performing fitting
routines to increase the quality and robustness of fits.
\end{abstract}

\section{Introduction}
Recent trends in commodity computing hardware have seen a dramatic
shift first from single-core processors to multi-core and then to
accelerated platforms like graphics processing units (GPUs). GPUs were
originally designed to speed up 3D graphics calculations for video
games, but their immense memory bandwidth and arithmetic capabilities
have seen them re-purposed for the needs of scientific
computing. While unquestionably powerful, their radically different,
massively-parallel architectures have shaken up the software
community. Astronomy is one of many fields trying to adapt to
these changes in computing hardware.

While the area is still in its infancy, GPUs have already been shown
to provide significant speed-ups across a range of astronomy
problems. These include direct N-body simulation (e.g., 
\citealt{HamadaEtal2009}), adaptive mesh refinement hydrodynamics (e.g.,
\citealt{SchiveEtal2010}), galaxy spectral energy density calculations
\citep{JonssonPrimack2010}, gravitational microlensing
\citep{BateEtal2010}, correlation for radio telescopes (e.g.,
\citealt{WaythEtal2009}) and coherent pulsar dedispersion
\citep{vanStratenBailes2010}. The approach taken in each of these
cases has, however, been \textit{ad hoc} in nature -- the transition
to the GPU has been guided largely by hardware-specific documentation,
code samples and simple trial and error. While such an approach has
proven very successful for these early adopters, it is not clear that
it will remain effective when it comes to more complex
algorithms. Furthermore, in some cases the cost of re-implementing a
code may be too large to gamble on a return (i.e., a speed-up) of
unknown magnitude.

In this paper we discuss the potential for accelerating the process of
galaxy fitting using GPUs. Rather than tackling the challenge blind,
we instead use a generalised method based on algorithm analysis as
outlined in \citet{BarsdellEtal2010}. The galaxy fitting process is
described in Section \ref{sec:GalaxyFitting}, which is followed by a
full analysis of the problem in Section \ref{sec:AlgorithmAnalysis}.
A preliminary implementation and results are described in Section
\ref{sec:Results} before our summary discussion in Section
\ref{sec:Discussion}.

\section{Galaxy Fitting}
\label{sec:GalaxyFitting}
A common problem in astronomy is to fit analytic surface brightness
profiles to observations of globular clusters or galaxies in order to
extract structural parameters such as the effective radius, ellipticity
or integrated flux magnitude. The fitting procedure is typically
non-linear and of high dimensionality, demanding the use of powerful
optimisation routines. Many codes have been developed to perform
this task, including, e.g.,
\textsc{ishape} \citep{Larsen1999},
\textsc{galfit} \citep{PengEtal2002} and
\textsc{galphat} \citep{YoonEtal2010}, which use the downhill simplex,
Levenberg Marquardt and Markov-Chain Monte-Carlo methods
respectively. While there is a variety of optimisation techniques in
common use, most follow a similar pattern:
\begin{enumerate}
\item Evaluate a model on a grid using the current set of parameters
  (guessed initially).
\item Convolve the model with the point spread function (PSF) of the observation.
\item Compare the model and observation.
\item Adjust the model parameters.
\item Check finishing criteria and return optimised parameters if
  complete, otherwise repeat from Step 1.
\end{enumerate}
Step 4 is where the specifics of a particular fitting routine come
into play, while steps 1-3 generally remain unchanged between
methods. Given the pixel-counts of modern astronomical observations,
and the fact that most fitting routines require a very large number of
iterations, the optimisation process can be highly computationally
intensive. The quantity and quality of science results are thus tied
to the available computing power and a code's ability to take
advantage of it.

Computationally-limited problems like galaxy fitting are ideal
candidates for acceleration. The fact that steps 1-3 are common to a
large number of fitting routines allows us to study the problem with
significant generality. Additionally, the image-based nature of the
operations immediately suggests suitability for GPUs.

\section{Algorithm Analysis}
\label{sec:AlgorithmAnalysis}
In order to determine whether galaxy fitting is a suitable application
for GPU acceleration, we use an approach based around algorithm
analysis as described in \citet{BarsdellEtal2010}. We begin by
identifying known algorithms within the steps in the problem outline
presented in Section \ref{sec:GalaxyFitting}:
\begin{enumerate}
\item Evaluation of a model on a grid is an example of a
  \textbf{transform} algorithm.
\item Convolution with the PSF is best done in Fourier space,
  requiring the \textbf{fast Fourier transform (FFT)} and regular
  \textbf{transform} algorithms.
\item Comparison of a model with an observation typically involves
  computation of the ``sum of squared differences'', which involves
  the \textbf{transform} and \textbf{reduce} algorithms.
\end{enumerate}
Note that the algorithms required during optimisation of the model
parameters in step 4 will depend on the chosen fitting routine.

It is thus seen that the fitting procedure makes use of only the
transform, FFT and reduce algorithms, all of which are known to be
very efficient on many-core architectures like GPUs
\citep{BarsdellEtal2010}.

The next step in the analysis is to look at the global characteristics
of the computation. Both the transform and reduce algorithms have a
work complexity of $O(N)$, indicating that a constant number of
operations is performed for every image pixel. The FFT algorithm
has a work complexity of $O(N\log{N})$, indicating that $O(\log{N})$
operations are performed for each of the $N$ image pixels. The
convolution step is thus expected to consume the majority of the
processing time for large images.

The fitting procedure for an image of $N$ pixels therefore requires
reading $O(N)$ values, repeatedly performing $O(N\log{N})$
arithmetic operations $O(N_{iter})$ times, and writing out $O(1)$
optimised parameter values (where $N_{iter}$ is the number of
iterations required to obtain a good fit). In the best case scenario
then, the problem has a ratio of memory to compute operations, or
\textit{arithmetic intensity}, of $O(N_{iter}\log{N})$.

The ability to achieve this arithmetic intensity depends on the memory
access patterns of the component algorithms. Because the FFT algorithm
requires an all-to-all communication pattern (i.e., each output value
depends on every input value), it is necessary to have the entire
image globally accessible during each iteration of the fitting
routine. This rules out storing the image data in very small caches
(e.g., the \textit{shared memory} on NVIDIA GPUs) between iterations.
However, there is generally more than enough main memory on a GPU to
hold an entire image. This means that the data may be left on the
device for all $N_{iter}$ iterations, with no need to go back 
to the host's memory or disk until the final results have been
obtained. The limiting factor will instead be the internal memory
bandwidth of the device. This is a good result, as current GPUs have
significantly more memory bandwidth than CPUs, and one can expect a
corresponding speed-up.

\section{Implementation Results}
\label{sec:Results}
Given the positive results of the algorithm analysis in Section
\ref{sec:AlgorithmAnalysis}, a prototype implementation of the galaxy
fitting problem was deemed worthwhile. NVIDIA's
CUDA\footnote{http://www.nvidia.com/object/cuda\_home\_new.html}
platform was used to interface to the GPU. FFTs were performed using
the CUFFT
library\footnote{http://developer.nvidia.com/object/cuda\_archive.html},
and the 
Thrust\footnote{http://code.google.com/p/thrust} C++ library was used
for its efficient implementations of the transform and reduce
algorithms.

Given the subtleties of mature codes like \textsc{galfit}
\citep{PengEtal2002}, performing an accurate comparison with our
prototype GPU code is not yet possible. Preliminary results, however,
suggest a speed-up in the main computations of around $10\times$ when
using a single NVIDIA Tesla C1060 GPU versus an Intel Nehalem
CPU. Profiling results also indicate that the GPU hardware is being
used efficiently by all of the algorithms in the code. These results
support the conclusions of our analysis in the previous section.

\section{Discussion}
\label{sec:Discussion}
Many-core architectures like GPUs are now an important part of the
computing landscape. While many software challenges remain, a
generalised approach to analysing astronomy problems has proven very
useful in tackling new GPU codes.

Galaxy fitting looks to be a promising application of GPU
technology. Significant speed-ups present the opportunity to perform
faster fits, which may be crucial for the next generation of galaxy
surveys. Alternatively, the additional processing speed could be fed
back into the fitting routine to provide fits of much better
quality in the same length of time, helping to overcome common
problems such as local minima and unphysical results.

While useful as a profiling tool, our prototype GPU code requires
significant further development before it can be considered a viable
alternative to other galaxy fitting codes in use by the astronomy
community. Future work will address such development.

Given the generality of our analysis, it is likely that other fitting
problems in astronomy would also benefit from GPU acceleration. If one
allows flexibility in the dimensionality of the problem, procedures
such as spectral line or cube fitting become possible. Such problems
will also be the subject of future work.

\bibliography{abbrevs,benbarsdell}

\begin{thebibliography}{}
\expandafter\ifx\csname natexlab\endcsname\relax\def\natexlab#1{#1}\fi
\expandafter\ifx\csname url\endcsname\relax
  \def\url#1{\texttt{#1}}\fi
\expandafter\ifx\csname urlprefix\endcsname\relax\def\urlprefix{URL }\fi
\providecommand{\eprint}[2][]{\url{#2}}

\bibitem[{{Barsdell} et~al.(2010){Barsdell}, {Barnes}, \&
  {Fluke}}]{BarsdellEtal2010}
{Barsdell}, B.~R., {Barnes}, D.~G., \& {Fluke}, C.~J. 2010, \mnras, 408, 1936.
  \eprint{1007.1660}

\bibitem[{{Bate} et~al.(2010){Bate}, {Fluke}, {Barsdell}, {Garsden}, \&
  {Lewis}}]{BateEtal2010}
{Bate}, N.~F., {Fluke}, C.~J., {Barsdell}, B.~R., {Garsden}, H., \& {Lewis},
  G.~F. 2010, New Astronomy. Accepted for publication June 2010,
  \eprint{1005.5198}

\bibitem[{{Hamada} et~al.(2009){Hamada}, {Nitadori}, {Benkrid}, {Ohno},
  {Morimoto}, {Masada}, {Shibata}, {Oguri}, \& {Taiji}}]{HamadaEtal2009}
{Hamada}, T., {Nitadori}, K., {Benkrid}, K., {Ohno}, Y., {Morimoto}, G.,
  {Masada}, T., {Shibata}, Y., {Oguri}, K., \& {Taiji}, M. 2009, Computer
  Science - Research and Development, 24, 21.
  \urlprefix\url{http://www.springerlink.com/content/j288l042547v4403}

\bibitem[{{Jonsson} \& {Primack}(2010)}]{JonssonPrimack2010}
{Jonsson}, P., \& {Primack}, J.~R. 2010, New Astronomy, 15, 509.
  \eprint{0907.3768}

\bibitem[{{Larsen}(1999)}]{Larsen1999}
{Larsen}, S.~S. 1999, \aaps, 139, 393. \eprint{arXiv:astro-ph/9907163}

\bibitem[{{Peng} et~al.(2002){Peng}, {Ho}, {Impey}, \& {Rix}}]{PengEtal2002}
{Peng}, C.~Y., {Ho}, L.~C., {Impey}, C.~D., \& {Rix}, H. 2002, AJ, 124, 266.
  \eprint{arXiv:astro-ph/0204182}

\bibitem[{{Schive} et~al.(2010){Schive}, {Tsai}, \& {Chiueh}}]{SchiveEtal2010}
{Schive}, H., {Tsai}, Y., \& {Chiueh}, T. 2010, ApJ.~Supp., 186, 457.
  \eprint{0907.3390}

\bibitem[{{van Straten} \& {Bailes}(2010)}]{vanStratenBailes2010}
{van Straten}, W., \& {Bailes}, M. 2010, ArXiv e-prints. \eprint{1008.3973}

\bibitem[{{Wayth} et~al.(2009){Wayth}, {Greenhill}, \&
  {Briggs}}]{WaythEtal2009}
{Wayth}, R.~B., {Greenhill}, L.~J., \& {Briggs}, F.~H. 2009,
  Pub.~Astron.~Soc.~Pacific, 121, 857. \eprint{0906.1887}

\bibitem[{{Yoon} et~al.(2010){Yoon}, {Weinberg}, \& {Katz}}]{YoonEtal2010}
{Yoon}, I., {Weinberg}, M., \& {Katz}, N. 2010, ArXiv e-prints.
  \eprint{1010.1266}

\end{thebibliography}

\end{document}